\documentstyle[preprint,aps]{revtex}                  %%%To be commented
\begin{document}
\pagestyle{empty}                                      %%%To be commented
\preprint{
\font\fortssbx=cmssbx10 scaled \magstep2
\hbox to \hsize{
\hbox{\hskip.2cm  hep-ph/yymmnn}
\hfill$\raise .5cm\vtop{\hbox{ NTUTH-95-09}
                \hbox{ NHCU-HEP-95-09}
                \hbox{}\hbox{}}$}
}
\draft
\vfill
\title{
Like Sign Top Quark Pair Production at Linear Colliders
}
\vfill
\author{Wei-Shu Hou}
\address{
Department of Physics, National Taiwan University,
Taipei, Taiwan 10764, R.O.C.
}
\author{Guey-Lin Lin}
\address{
Institute of Physics, National Chiao Tung University,
Hsinchu, Taiwan 30050, R.O.C.
}
\date{\today}
%
%\vskip -1cm
%
\vfill
\maketitle
\begin{abstract}
In the general two Higgs doublet model with
flavor changing neutral Higgs couplings,
neutral Higgs bosons may decay dominantly via
$t\bar c$ or $\bar tc$ final states.
At the linear collider, $e^+e^- \to h^0A^0$ or
$H^0A^0$ production processes may result in
$b\bar bt\bar c$, $W^+W^-t\bar c$ or
$tt\bar c\bar c$ (or $\bar t\bar tcc$)  final states.
The process $\gamma\gamma \to h^0,\ A^0 \to t\bar c$ is also promising,
but $e^+e^- \to (h^0,\ H^0) Z^0 \to t\bar c Z^0$
is relatively suppressed.
The possibility of observing like sign lepton pairs,
usually the hallmark for neutral meson mixing,
is quite interesting since  $T^0$ mesons do not even form.
\end{abstract}
%
%\vfill
%
\pacs{PACS numbers:
14.80.Dq, %Specific Properties of Quarks
14.80.Gt, %Higgs Boson
12.15.Cc, %extended Higgs (gauge) sector
13.90.+i  %other
}
%
%\narrowtext
%
\pagestyle{plain}

%\section{Introduction}

Like sign dilepton pair production is the hallmark
for  heavy neutral meson--antimeson mixing.
The standard model (SM) predicts rather small mixing effects
for mesons containing $u$-type quarks.
Furthermore, due to its heaviness,
the top quark decays before the  $T_u^0$ or $T_c^0$ mesons could form.
Thus, unlike the $b$ quark case,
we do not expect {\it same} sign dileptons
from $t\bar t$ pair production.
Effects beyond the standard model are
not expected to change this,
since the Tevatron data \cite{CDF} is in good agreement with
$t\to bW$ decay dominance expected in SM.
In this note we report \cite{LC95}
the  intriguing possibility of producing like sign top quark pairs
at linear $e^+e^-$ colliders,
within the context of a general two Higgs doublet model (2HDM)
that possesses flavor changing neutral Higgs (FCNH) couplings
\cite{CS,Hou,LS}.

Atwood, Reina and Soni have recently studied \cite{ARS} FCNH loop induced
$e^+e^- \to \gamma^*,\ Z^* \to t\bar c$ transitions
at linear colliders and find a rather small rate.
They also propose \cite{ARS2}
to study the tree level $s$-channel FCNH process
$\mu^+\mu^- \to \mbox{neutral scalars} \to t\bar c$.
Here, we explore FCNH coupling effects
in Higgs boson production processes at
a 500 GeV $e^+e^-$ Next Linear Collider (NLC).
We find it to be promising,
both for single top $t\bar c + X$,
as well as for the more intriguing
like sign top pair $tt\bar c\bar c$ final states.

%\section{Model and Constraints}

Let us briefly review the model under consideration.
With two Higgs doublets $\Phi_1$ and $\Phi_2$,
in general one has FCNH couplings.
Because of stringent bounds from $\mu\to e\gamma$ decay,
$K^0$--$\bar K^0$ and $B^0$--$\bar B^0$ mixings, {\it etc.},
it is customary \cite{Hguide} to strictly enforce
the absence of FCNH couplings at tree level.
This is readily achieved via some discrete
symmetry that allows just one source of mass
for each given fermion charge \cite{GW}, much like in SM.
However, inspired by the quark mass and mixing hierarchy pattern
\begin{equation}
   \begin{array}{ccccccc}
    m_1 & \ll & m_2 & \ll & m_3, && \\
    \vert V_{ub}\vert^2 & \ll & \vert V_{cb} \vert^2 &
                      \ll & \vert V_{us}\vert^2 & \ll &1,
   \end{array}
\end{equation}
that emerged since the early 1980's,
Cheng and Sher \cite{CS} suggested that
low energy flavor changing neutral currents
could be naturally suppressed,
without the need to invoke discrete symmetries.
Let us elaborate on this observation.

We shall assume $CP$ invariance throughout the paper,
leaving out even the possibility of spontaneous $CP$ violation \cite{WW}.
Since both  $\Phi_1$ and $\Phi_2$ develop real
vacuum expectation values ({\it v.e.v.}),
one can redefine the fields and choose
one doublet as the ``mass giver", {\it i.e.}
$\langle\phi_1^0\rangle = v/\sqrt{2}$,
$\langle\phi_2^0\rangle = 0$,
where $v \simeq 246$ GeV.
The usual 2HDM parameter $\tan\beta\equiv v_1/v_2$
gets rotated away by the freedom to make linear redefinitions.
One readily sees that \cite{LS}
$\sqrt{2}\, \mbox{Re}\,\phi_1^0$ has diagonal couplings.
It is, however, {\it not} a mass eigenstate.
For $\Phi_2$ related fields, we have,
\begin{eqnarray}
   \left(\bar u_{L} \xi^{\left(u\right)} u_{R}
               + \bar d_{L} \xi^{\left(d\right)} d_{R}\right)\,
                      \sqrt{2}\, \mbox{Re}\, \phi_2^0
      &\ + \ & \left(-\bar u_{L} \xi^{\left(u\right)} u_{R}
               + \bar d_{L} \xi^{\left(d\right)} d_{R}\right)\,
                 i\,\sqrt{2}\, \mbox{Im}\, \phi_2^0 \nonumber \\
   - \; \bar d_{L} V^\dagger \xi^{\left(u\right)} u_{R}\,  \sqrt{2}\, \phi_2^-
       &\ + \ & \bar u_{L} V\,       \xi^{\left(d\right)} d_{R}\, \sqrt{2}\,
\phi_2^+
                \ + \mbox{\ H.c.},
\end{eqnarray}
where $\xi^{\left(u,d\right)}$ are in general
{\it not} diagonal, but
$V^{\left(\dagger\right)}\xi \simeq \xi$,
since the KM matrix $V \simeq 1$.

At first sight, the Yukawa coupling matrices $\xi^{\left(u,d\right)}$ may
appear to be
completely general.
However, in some arbitrary basis
where $\langle\phi_1^0\rangle = {v_1/\sqrt{2}}$,
$\langle\phi_2^0\rangle = v_2/\sqrt{2}$,
quark mass matrices consists of two parts,
$m = m^{\left(1\right)} + m^{\left(2\right)}$.
To sustain eq. (1),
unless fine-tuned cancellations are implemented,
one would expect that the off diagonal elements
of $m^{\left(1\right)}$ {\it and} $m^{\left(2\right)}$, just like $m$ itself,
should trickle off as
one moves off-diagonal.
The rotation (linear redefinition) by angle $\beta = \tan^{-1}(v_1/v_2)$
to eq. (2) should not change this property.
Hence, {\it data} (eq. (1)) {\it suggest that $\xi$ cannot be arbitrary}.
In this vein, Cheng and Sher proposed \cite{CS}
the ansatz
\begin{equation}
\xi_{ij} \sim {\sqrt{m_i m_j}/v}.
\end{equation}
The bonus was that FCNH couplings involving
lower generation fermions are naturally suppressed,
without the need to push FCNH Higgs boson masses
to way beyond the {\it v.e.v.} scale \cite{McWLS}.
Inspecting eq. (1) again,
a weaker ansatz is possible \cite{LC95},
\begin{equation}
\xi_{ij} = {\cal O}(V_{i3}V_{j3})\, m_3/v.
\end{equation}
According to the mass--mixing pattern,
the Cheng-Sher ansatz of eq. (3) corresponds to
$\xi_{ij} = {\cal O}(V_{ij}V_{j3})\, m_3/v$.
Note that in both cases, $\xi_{ct}$ is the largest possible FCNH coupling,
and the associated phenomenology is the most interesting \cite{Hou,HW}.

The pseudoscalar $A^0 \equiv \sqrt{2}\, \mbox{Im}\, \phi_2^0$
and charged scalar $H^\pm \equiv \phi_2^\pm$ are already physical
Higgs bosons, but the neutral $CP$ even Higgs bosons
$H^0$ and $h^0$ are mixtures of
$\sqrt{2}\, \mbox{Re}\,\phi_1^0$ and $\sqrt{2}\, \mbox{Re}\,\phi_2^0$.
The mixing angle $\sin\alpha$, a physical parameter,
is determined by the Higgs potential.
In the limit of $\sin\alpha \to 0$,
often assumed by various authors \cite{LS,ARS,ARS2},
$H^0 \leadsto \sqrt{2}\, \mbox{Re}\, \phi_1^0$
becomes the ``standard" Higgs boson with
diagonal couplings,
while
$h^0 \leadsto \sqrt{2}\, \mbox{Re}\, \phi_2^0$
has Yukawa couplings as in eqs. (3) or (4),
but decouples from vector bosons
or charged Higgs bosons, just like $A^0$.
Our convention for $H^0$ and $h^0$
differs from the minimal supersymmetric SM (MSSM) \cite{Hguide},
where $h^0$ is taken as the lighter $CP$ even neutral scalar.

Constraints on the general 2HDM has been studied by various authors.
For $K^0$ and $B^0$ mixings,
one finds \cite{CS,SY} a rather weak bound of $m_{h^0} \gtrsim 80$ GeV,
with a more stringent bound for $A^0$.
These bounds could weaken, for example,
if one uses eq. (4) instead of eq. (3).
For $\mu \to e\gamma$, an interesting two loop effect
dominates over one loop diagrams \cite{CHK}.
{}From Fig. 4 of ref. \cite{CHK}, with $m_t \simeq 175$ GeV,
one finds a bound of $m_{h^0} \gtrsim 150$ GeV.
The bound for $A^0$ is weaker since it does not couple to vector bosons
and unphysical scalar bosons.
If $\sin\alpha \to 0$,
the $h^0$ bound would also weaken.
A third, less direct constraint on FCNH Higgs boson masses
is from the recent experimental observation of
inclusive $b\to s\gamma$ decays.
Within the the so-called Model II of 2HDM \cite{Hguide}
(automatically realized in MSSM),
the CLEO Collaboration gives \cite{CLEO} a bound of $m_{H^+} > 250$ GeV.
This bound should weaken for our case because of the freedom in
$\xi^{\left(u,d\right)}$ as compared to Model II.
Inclusion of next-to-leading order
QCD corrections also tends to soften the bound \cite{Ciu}.
Thus, we take $m_{H^+} \gtrsim 150 - 250$ GeV
as  a reasonable lower bound, which is rather consistent with
the bounds on FCNH neutral scalar bosons.
The upshot of our discussion on low energy constraints is that,
\begin{equation}
v\sim m(\mbox{FCNH Higgs}) \gtrsim m_t
\end{equation}
is not only reasonable, but quite likely.
Although $t\to c\ +$  scalar transitions \cite{Hou,HW} are not excluded,
we are more interested in Higgs bosons
decaying into $t\bar c$ \cite{Hou}.

%\section{Production and Decay}

We will focus mainly on the mass domain of
\begin{equation}
200\ \mbox{GeV} < m_{h^0,\; A^0} < 2m_t \simeq 350\ \mbox{GeV}.
\end{equation}
%The upperbound of 350 GeV also roughly reflects the reach of the NLC,
%while the lower range of 200 GeV allows $h^0,\ A^0 \to t\bar c$ decay
%to be suitably above threshold.
We plot in Fig. 1 $BR(S^0 \to t\bar c + \bar tc)$
{\it vs.} $\sin^2\alpha$, for $S = h$, $A$ and $H$
and $m_{S^0} = 200$, $250$, $300$ GeV.
$A^0$ can decay only via
$t\bar c$ and $f\bar f$ modes, and can be treated as
independent of $\sin^2\alpha$.
Once $m_{A^0}$ is suitably above $tc$ threshold,
the $t\bar c$ or $\bar tc$ modes dominate.
The behavior for $h^0$ at $\sin\alpha = 0$ is similar to $A^0$.
However, as $\sin^2\alpha$ grows,
the $h^0 \to V\bar V$ ($V = W,\ Z$) partial width grows rapidly,
and $t\bar c$ branching ratio becomes rather suppressed.
The case for $H^0$ is the same as $h^0$ under the interchange
of $\sin^2\alpha \longleftrightarrow \cos^2\alpha$.
The proximity of the $m_{h^0,H^0} = 250$ and $300$ GeV curves is accidental.

One might think that the most promising channel for studying
FCNH Higgs bosons is via the associated production process of
$e^+e^- \rightarrow  Z^* \rightarrow H^0Z^0$ and $h^0Z^0$
(The $W^+W^-$ fusion process $e^+e^-  \rightarrow  \nu\bar\nu + H^0$
is subdominant for the range of eq. (6)).
This turns out to be not the case.
We plot in Fig. 2 the cross section times branching ratio for
the signature $e^+e^- \to  S^0Z^0 \to t\bar cZ^0$
{\it vs.} $\sin^2\alpha$, where $S = H,\ h$,
again for $m_{S^0} = 200,\ 250,\ 300$ GeV.
For $H^0$ this behaves as
$x(1-x)/(A(1-x) + ax + \delta)$ where $x \equiv \sin^2\alpha$,
$A$ is related to the $H\to V\bar V$ rate in SM,
and $a$, $\delta$ are related to the $t\bar c$ and $b\bar b$ rate.
For $h^0$ case one just interchanges $x \rightarrow 1-x$.
Clearly, in the limit of  $\sin\alpha = 0$,
$H^0$ couples only flavor diagonally,
while $h^0$ has no production cross section.
As $\sin\alpha$ grows,
because of the mismatch in the production and decay process,
the effective cross section
for $t\bar c + Z^0$ associated production
remains rather small, with the maximum of 0.43 fb
at $m_{h,H} \cong 237$ GeV and
$\sin^2\alpha$ $(1-\sin^2\alpha$ for $H^0) \cong 0.129$.
This is but a fraction of the total $e^+e^- \to  S^0Z^0$ cross section,
which would not be easy to observe once
one folds in various branching ratios for $t$ or $Z$ decay.

What is more promising is the
$e^+e^- \rightarrow Z^* \rightarrow S^0A^0$
associated production process, where $S = h,\ H$.
In the $\sin\alpha \to 0$ limit,
one has $e^+e^- \to h^0A^0$ only,
with cross section similar to the $H^0Z^0$ mode when
phase space is similar.
Since in this limit,
$h^0 \not\to V\bar V$,
the $t\bar c$ mode has a good chance to be
the dominant final state for both $h^0$ and $A^0$.
We immediately see the possibility of our purported
$tt\bar c\bar c$ or $\bar t\bar tcc$ final states!

Let us proceed a bit more systematically.
Taking $m_{A^0} > m_{h^0}$ for illustration
(since $A^0\to t\bar c$ is quite likely to be dominant), we allow
$m_{h^0}$ to be as low as 100 GeV,
with $m_{A^0}$ in the range of eq. (6),
but $A^0 \to h^0 Z^0$ is kinematically suppressed or forbidden.
For numerical illustration, we give in Table \ref{table1}
the number of $h^0A^0$ events at an
NLC with 50 fb$^{-1}$ integrated luminosity,
for $\sin\alpha = 0$, $m_A^0 = 200,\ 250$ GeV,
$m_{h^0} = 100,\ 150,\ 200$ GeV,
with the condition that $m_{A^0} - m_{h^0} < 100$ GeV.
We see that, up to phase space,
a few$\; \times 10^2$ to $10^3$ raw events are expected.
To determine the number of signal events,
one could easily fold in branching ratios from Fig. 1,
and $\cos^2\alpha$ or $\sin^2\alpha$ factors
for the production cross section.
These should then be compared with potential backgrounds,
the chief ones being $e^+e^- \to t\bar t,\ W^+W^-,\ Z^0Z^0$ events,
which are of order $3\times 10^4$, $4\times 10^5$, $3\times 10^4$,
respectively.
Assuming  $A^0 \to t\bar c + \bar tc$ is predominant,
three modes are of interest:
$h^0\to b\bar b,\ t\bar c$ and $V\bar V$.

For $m_{h^0} < 2M_W$,
or when $\sin^2\alpha \ll 0.1$
but the $t\bar c$ mode is suppressed or forbidden,
$h^0 \to b\bar b$ would be
the dominant decay mode.
We have $\sigma(e^+e^- \to h^0A^0) \times
 \mbox{BR}(h^0\to b\bar b)\mbox{BR}(A^0\to t\bar c + \bar tc)
\sim 10 - 20$ fb, which
is quite sizable compared to the $t\bar cZ^0$ case.
However, the $b\bar b t\bar c$ final state may
take some effort to identify,
since there is just one top quark.
Although kinematic tricks could be played,
but faced with backgrounds that are
orders of magnitude higher,
one would need very good $b$-tagging efficiency,
and would likely need to know $m_{h^0}$ beforehand.
However, the latter may have to be studied at the NLC itself,
unless the intermediate mass Higgs search program
(via $H \to \gamma\gamma$ detection) at the LHC turns out to be
very successful.
In any case a detailed Monte Carlo study would be necessary
to determine whether this mode can be fruitfully studied.

For $\sin^2\alpha \ll 0.1$,
as can be seen from Fig. 1, $h\to t\bar c$ is likely dominant
in the mass range of eq. (6).
We find
$\sigma(e^+e^- \to h^0A^0) \times
 \mbox{BR}(h^0\to t\bar c + \bar tc)\mbox{BR}(A^0\to t\bar c + \bar tc)
\sim 10 $ fb,
which is slightly smaller than the previous case because of phase space.
Clearly, $50\%$ of this cross section
goes into $tt\bar c\bar c$ or $\bar t\bar tcc$ final states,
which is again larger than the $t\bar cZ^0$ case.
We expect typically of order 250 such events.
Folding in the semileptonic branching ratio,
one expects $\sim 12$ events in the signal of
\begin{equation}
e^+e^- \longrightarrow \ \ \ell^\pm\ell^{\prime\pm} + \nu\nu + 4j,
\end{equation}
where the 4 jets have flavor $bb\bar c\bar c$ or $\bar b\bar bcc$.
Thanks to the large top quark mass,
this distinctive signature has seemingly {\it no background}.
In contrast, if one allows only one top to decay semileptonically,
or if one tries to probe the equivalent number of $t\bar tc\bar c$ events,
the single $\ell + \nu + 6j$ or
opposite sign dilepton $\ell^\pm\ell^{\prime\mp} + \nu\nu + 4j$
signatures would be swamped by $t\bar t$ or $W^+W^-$ production
background, which are orders of magnitude higher.
In particular,
standard $e^+e^- \to t\bar t$ pair production
with hard gluon radiation may be especially irremovable.
Since the effect demands $m_{h^0} + m_{A^0} > 400$ GeV,
one is phase space limited at a 500 GeV NLC.
If the center of mass energy of the NLC could be increased
to 600 GeV or so, possible phase space suppressions for
producing $e^+e^- \to h^0A^0 \to tt\bar c\bar c + \bar t\bar tcc$
could be relieved.

For $\sin^2\alpha \gtrsim 0.1$ and $m_{h^0} > 2M_W$,
$h^0 \to W^+W^-$ and $Z^0Z^0$ decays are (pre)dominant,
with $h^0 \to t\bar c$ no more than $10\%$,
and like sign top quark pair final states
become no longer visible.
The case remains interesting, however,
since $W^+W^-t\bar c$ or $Z^0Z^0t\bar c$ final states
at the $5-10$ fb level (taking into account both $h^0$ and $H^0$)
are still quite conspicuous.
One effectively has $W^+W^-W^+b\bar c$ in final state,
which again has little background.
In particular, one could still have like sign dilepton pairs as in
eq. (7).
With $m_{h^0}$ or $m_{H^0}$ known from the LHC,
this decay mode could be studied
at the NLC in complete detail.

%\section{Discussion and Summary}

It is also possible to produce Higgs boson via
the $\gamma\gamma\to S^0$ process \cite{Hguide}.
We note that
$A^0$ only couples to fermions, hence its
effective coupling to photons is smaller than the SM Higgs boson.
For $h^0$, if $\sin\alpha$ is very small, the case is again
similar.
As $\sin\alpha$ grows, the effective coupling
would quickly become dominated by vector bosons
and the production cross section could be larger.
This is, however, offset by the reduction in $h^0 \to t\bar c$
branching ratio.
Since  $\gamma\gamma\to S^0 \to t\bar c$
should have little background
({\it e.g.} $t\bar c \to W^+ b\bar c$ can be distinguished from
$W^+W^-$ via $b$-tagging since BR$(W^- \to b\bar q) < 10^{-3}$),
the number of events expected is \cite{GH}
\begin{eqnarray}
   N(\gamma\gamma\to S^0 \to t\bar c + \bar tc)
&=& 4\pi^2 \Gamma(S^0 \to \gamma\gamma)\,
   \mbox{BR}(S^0\to t\bar c + \bar tc)/m_{S^0}^2 \nonumber\\
&\times & (1 + \langle\lambda\lambda^\prime\rangle)
        \, \langle d{\cal L}_{\gamma\gamma}/dm_{\gamma\gamma}\rangle
                                     \vert_{m_{\gamma\gamma} = m_{S^0}}.
\end{eqnarray}
It is possible \cite{gamgam} to tune photon polarizations to
have $\langle\lambda\lambda^\prime\rangle \sim +1$
and effective luminosities close to the the $e^+e^-$ mode
({\it i.e.}  $\sim 50$ fb$^{-1}$).
If such is the case, then one    %typically
expects $10^2 - 10^3$ raw events,
which should make $t\bar c$ detection possible
if the branching ratio is not too suppressed.
Note that the corresponding number of
$\gamma\gamma \to W^+W^-$ pairs is at the $10^4 - 10^5$ level.

We now compare our results with that of Atwood, Reina and Soni.
For $e^+e^- \to \gamma^*,\; Z^* \to t\bar c$ via FCNH loop effects,
they find $R^{tc} \equiv
\sigma(e^+e^- \to t\bar c + \bar tc)/\sigma(e^+e^- \to \gamma^* \to \mu^+\mu^-)
\lesssim \mbox{few} \times 10^{-5}$ \cite{ARS},
which amounts to less than $0.1$ event for a 500 GeV NLC
with 50 fb$^{-1}$ integrated luminosity.
In case $h^0$ and $A^0$ are heavier than the range of eq. (6),
the loop induced cross section also
goes down by another order of magnitude \cite{ARS}.
Thus, this process is unlikely to be observable at the NLC.
Although phase space favored,
loop suppression in this case is too severe.
For $\mu^+\mu^- \to h^0$, $A^0 \to t\bar c + \bar tc$,
the process occurs at tree level
and has a sizable cross section \cite{ARS2}.
But in the limit of $\sin\alpha \to 0$,
$h^0$ would also not decay via the $V\bar V$ mode,
just like $A^0$.
A rather fine-stepped energy scan would then be needed
because of the narrowness of the $h^0$ and $A^0$ width.
Together with the technological uncertainty for a high energy,
high luminosity $\mu^+\mu^-$ collider \cite{mumu},
this process might be less straightforward to study
than at the NLC,
including the $\gamma\gamma$ collider option
via $\gamma\gamma \to h^0,\ A^0$.

As stated in the Introduction,
the signature of like sign top pair production
is rather analogous to observing
$T^0T^0$ or $\bar T^0\bar T^0$ pairs
via $T^0$--$\bar T^0$ mixing.
With top mesons not even forming, however,
it is the associated production of $h^0A^0$ pairs,
which each subsequently decay equally into
$t\bar c$ or $\bar tc$ final states,
that circumvents the usual condition of associated production
of $t\bar t$ (or, $W^+W^-$) pairs in most processes.
Since $h^0$ and $A^0$ contribute to $B$--$\bar B$ mixing,
in a sense the like sign top pair production effect
is related to neutral meson mixing phenomena.
We know of no other way to make $tt$
or $\bar t\bar t$ pairs in an $e^+e^-$ collider environment.

In summary, within a general two Higgs doublet model with
FCNH couplings,
$h^0$, $A^0 \rightarrow t\bar c + \bar tc$ could be the
dominant decay mode.
The most intriguing consequence is the possibility of
detecting like sign top pair production
via $e^+e^-\to h^0A^0 \to tt\bar c\bar c$ or $\bar t\bar tcc$,
while single top $t\bar cb\bar b$ or
$t\bar cW^+W^-$ production are also detectable.
In contrast, the $e^+e^- \to (H^0,\; Z^0) \to t\bar cZ^0$ process
is rather suppressed and not competitive.
The number of events, hence the FCNH Higgs boson mass reach,
could be extended if the collider energy is higher.
It is also possible to study single FCNH Higgs production
via $\gamma \gamma \to h^0,\ A^0 \to t\bar c + \bar tc$.
Since the neutral Higgs bosons of minimal supersymmetric standard model
couple to fermions in a flavor diagonal way, the
{\it observation of FCNH signals
would rule out MSSM}.
We urge experimental colleagues to
study signal {\it vs.} background issues carefully.

\acknowledgments
The work of WSH is supported in part by grant NSC 85-2112-M-002-011,
and GLL by grant NSC 85-2212-M-009-006
of the Republic of China.

 \begin{figure}
 \caption{$BR(S^0 \to t\bar c + \bar tc)$
  {\it vs.} $\sin^2\alpha$ for $S = h$ (dash), $A$ (dotdash) and $H$ (solid)
  and $m_{S^0} = 200$, $250$, $300$ GeV (from bottom to top).}
 \end{figure}

 \begin{figure}
 \caption{$\sigma(e^+e^- \to S^0Z^0) \times$
  BR$(S^0\to t\bar c)$ {\it vs.} $\sin^2\alpha$,
  for $S = h$ (dash), $H$ (solid)
  and $m_{S^0} = 200$ (lower curve), $250$ (upper curve), $300$
  (middle curve) GeV.}
 \end{figure}

 \begin{table}
 \caption{Number of $e^+e^- \to h^0A^0$ events
  for $\sin\alpha = 0$ at NLC with $\int {\cal L}\, dt = 50$ fb$^{-1}$.}
 \begin{tabular}{lrrrrrd}
  $m_{A^0}$ (GeV)     &   200 &  200 & 250 & 200 & 250     \\ \tableline
  $m_{h^0}$ (GeV)      &   100 &  150 & 150 & 200 & 200     \\ \tableline
  ${\cal N}(h^0 A^0)$  & 1160 & 900 & 490 & 520 & 200     \\
 \end{tabular}
 \label{table1}
 \end{table}

\end{document}